# Measuring algorithmic complexity in chaotic lasers

Marcelo G. Kovalsky, Mónica B. Agüero, Carlos Bonazzola and Alejandro A. Hnilo
*CEILAP, Centro de Investigaciones en Láseres y Aplicaciones, UNIDEF (MINDEF-CONICET);*
*CITEDEF, J.B. de La Salle 4397, (1603) Villa Martelli, Argentina.*
*Email: mkovalsky@citedef.gob.ar*
February 14th, 2019.

Thanks to the simplicity and robustness of its calculation methods, algorithmic (or Kolmogorov) complexity appears as a useful tool to reveal chaotic dynamics when experimental time series are too short and noisy to apply Takens' reconstruction theorem. We measure the complexity in chaotic regimes, with and without extreme events (sometimes called optical rogue waves), of three different all-solid-state lasers: Kerr lens mode locking femtosecond Ti: Sapphire ("fast" saturable absorber), Nd:YVO4 + Cr:YAG ("slow" saturable absorber) and Nd:YVO4 with modulated losses. We discuss how complexity characterizes the dynamics in an understandable way in all cases, and how it provides a correction factor to the horizon of predictability given by Lyapunov exponents. This approach may be especially convenient to implement schemes of chaos control in real time.

## 1. Introduction.

Chaotic dynamics of lasers has been intensively investigated both theoretical and experimentally. A standard approach to describe a complex nonlinear system is based on solving numerical simulations of the full set of differential equations. Other approaches study an associated discrete time system, the iterative, stroboscopic or Poincaré map. The experimental approach attempts to reconstruct the attractor from time series with appropriate time delay [1] and it relies mostly on the embedding theorem due to Takens and Mane [2]. These calculations are difficult and time-consuming. They are impractical to determine on the spot (f.ex., for chaos control purposes), if the laser behavior is complex, or simply noise.

On the other hand, Kolmogorov [3], Chaitin [4] and Solomonoff [5] defined a powerful tool to distinguish randomness from chaos. Kolmogorov defines *complexity* of a finite binary string $N$ as the length $K$ of the shortest binary program which, when run in some Universal Turing Machine, causes that machine to output the $N$ string and then halt. A sequence is "algorithmically random" if $K \approx N$. There is no way of expressing the sequence using less bits that the sequence itself, so that the sequence is said to be *incompressible*. This definition is intuitive and appealing, but it has two main drawbacks. One: it is possible to demonstrate that all series are (partially) compressible; hence, the precise condition $K \approx N$ cannot be reached. In the practice this problem is solved by appropriately rescaling the definition. Two: $K$ cannot be actually computed, for one can never be sure that there is no shorter program able to generate the sequence.

The reconstruction methods involve an a priori hypothesis (i.e., that there is an underlying attractor). Besides, it is not always possible to obtain the embedding dimension (dE) from an experimental time series. The reason, in a particular time series, is often that the system has been observed in a small volume only, so sampling is too scarce to reconstruct the underlying attractor. This kind of series is called *noisy* in the jargon. Instead, Kolmogorov complexity makes no assumption on the origin of the time series.

Recently, Kolmogorov complexity was applied to reveal chaotic and random behaviors from measurements in fields such as quantum-based random number generators [6], earth [7] and computer sciences [8]. Surprisingly, this powerful tool has not been applied to study nonlinear laser dynamics. In this paper, we compute the algorithmic complexity of time series generated in chaotic lasers by using the realization of the Lempel and Ziv [9] algorithm (LZA) developed by Kaspar and Schuster [10]. In particular, a relevant behavior of some chaotic regimes is the occurrence of extreme events (EE). In lasers, an EE is defined as a pulse whose amplitude is larger than a certain threshold. This threshold has often been defined as twice the average value of the amplitude of the largest one-third maxima in the total distribution [11]. It is recognized that general EE may have different causes. They are observed in fields such different as hydrodynamics [12], nonlinear fiber optics [13] and Bose Einstein condensates [14]. Here we show that Kolmogorov complexity is an accurate tool to detect chaos in three different chaotic lasers (with and without EE) even if working with series with a large amount of noise. We also show how it can be used to correct the usual horizon of predictability given by the maximum Lyapunov exponent. Section 2 describes the fundamentals of the algorithm as well as its implementation. Section 3 defines *predictability* in practical terms. Section 4 is devoted to present the results obtained in three different lasers: Kerr lens mode locking (KLM) femtosecond Ti:Sapphire (laser + fast saturable absorber), diode pumped Nd:YVO4 with Cr:YAG (laser + slow saturable absorber) and Nd:YVO4 with active modulation of the losses

## 2. Algormith Complexity.

The LZA is a code for lossless data compression. It is actually not a single code, but a whole family of codes, stemming from the one proposed by Jacob Ziv and Abraham Lempel in their landmark paper in 1976 [9]. LZA are widely used in compression utilities such as gzip and GIF image compression standards [15]. The original LZA proposes to evaluate the complexity of a finite sequence from the point of view of a simple self-delimiting learning machine which, as it scans a given $n$-digit sequence S = {$s_1$, $s_2$ ... $s_n$} from left to right, adds a new word to its memory every time it discovers a substring of consecutive digits not previously encountered.

The size of the compiled vocabulary, and the rate at which new words are encountered along the sequence, serve as the basic ingredients in the evaluation of the complexity of the sequence. In the practical realization of the algorithm, the time series is encoded by constructing a sequence consisting of the characters 0 (1), if each element is smaller (larger) than some threshold. The mean value of the series is often used as that threshold [21]. In this way a balanced binary string is produced. Then the complexity counter $c(N)$, which is defined as the minimum number of distinct words in a given sequence of length $N$, is calculated. As $N \to \infty$, $c(N) \to N/log_2(N)$ in a random series. The normalized complexity measure $KC$ is then defined as:

$$KC(N) \equiv c(N) \times log_2(N)/N \qquad (1)$$

The value of $KC(N)$ is near to 0 for a periodic or regular time series, and 1 for a random one, assuming the value of $N$ is large enough. For a chaotic series it is typically between 0 and 1.

### 3. Predictability of time series.

One of the most remarkable results of the study of (chaotic) dynamical systems is that, although the evolution is ruled by deterministic equations, its state can be unpredictable beyond some point. This leads to the idea of a *prediction horizon*, usually estimated as $H \equiv 1/\lambda_{max}$ where $\lambda_{max}$ is the largest positive Lyapunov exponent. The Lyapunov exponent is a global parameter that measures the sensibility to initial conditions, based on the nonlinear interactions of few degrees of freedom. In this analysis noise or intrinsic randomness (relevant when experimental time series are studied) is not taken into account. Hence, $H$ is, in general, an overestimation of the actual value of predictable time ahead. To correct this effect, the *Kolmogorov time* $K_T \equiv 1/KC$ (in units of the recorded series) has been introduced [16]. It is estimated to be proportional to the randomness of the series, and quantifies the size of the time window within which complexity remains unchanged. In this way, the *effective* prediction horizon is shortened in presence of a narrow $K_T$ window.

Determinism and randomness seem opposite concepts. Yet, when dealing with finite experimental time series, they can be seen as the extremes of the same property: partial determinism. As a practical criterion, an experimental time series is considered random beyond the effective prediction horizon.

### 4. Results.

Time series of lasers' outputs are recorded with a fast pin photodiode (100ps risetime) connected to a PC oscilloscope (PicoScope6403B, 500-MHz bandwidth, 5 GS/s, memory 1 GS). From this raw material we can build two types of time series: one from the peak pulse intensities between consecutive pulses and other from the time separation between consecutive pulses. In the Ti:Sapphire laser, we consider peak pulse intensities only, because the peak to peak time is fixed by the cavity length.

*4. a. KLM Ti:Saphhire laser (fast saturable absorber).*

This laser is the most widespread source of ultrashort pulses nowadays. It is used in a number of applications, both academic and commercial. Self-pulsing is the consequence of a Kerr nonlinearity in the active medium, which can be considered as a fast saturable absorber. Our laser cavity is the standard seven-element design, (Fig.1) which uses a pair of prisms to introduce negative group-velocity dispersion (GVD) as the main control parameter. Our prototype emits trains of light pulses of duration between 20 and 200 fs at a rate of 87MHz. Two coexistent pulsed modes of operation exist, named P1 (transform limited output pulses) and P2 (chirped output pulses).[17] They are easily recognizable in the experiment, and were described in detail in [18-19]. The laser spontaneously passes from one mode to the other. A remarkable feature is that P2 displays EE, while P1 does not. [20]. We calculate the Kolmogorov complexity for several time series both from P1 and P2 (Table 1).

In all cases $KC$ is able to distinguish periodic behavior ($KC$ close to 0, as when $P_1\Delta(-150\ fs^2)$) from chaotic one ($KC$ intermediate between 0 and 1), even for noisy time series like $P_1\Delta(-37fs^2)$, where it is impossible to calculate the embedding dimension and hence the Lyapunov exponent. Note that in this case $K_T$ is the only prediction horizon available. In the coexistence zone between transform limited pulses and chirped pulses, $P_1 - P_2 \Delta(-40\ fs^2)$ as well as in the $P_1\Delta(-90\ fs^2)$ time series, the window given by $K_T$ almost coincides with the inverse of the largest Lyapunov exponent. Hence, the effective prediction horizon is unaffected by intrinsic randomness or noise in this case.

EE in this laser arise from a modulation instability-like mechanism. Nevertheless, the periodical orbits of the "cold cavity" survive as a quasi-periodicity in the separation between two consecutive EE. Measured in number of round trips, this separation turns out to be a simple combination of the numbers 11 and 12 (in our setup). In other words, we know that the next EE will arrive after a number of round trips given by:

$$\Delta \tau = 11 \times n + 12 \times m \qquad n, m \in \mathbb{N} \qquad (2)$$

This does not imply complete predictability, because we do not know the time of the first EE in the series or the values of $n$ and $m$. Yet, it means an increase in the information we have about the system. We calculate $KC$ of two time series, with and without EE, $P_2\Delta(-27\ fs^2)$ and $P_2\Delta(-32\ fs^2)$. There is a sudden drop from 0.5280 to 0.1289, that reflects the regularity in $\Delta\tau$ or, in other words, that the time series with EE is more compressible than the one without EE. The maximum Lyapunov exponent also reflects this regularity, as a larger value of $H$. Combining the information from $K_T$ and $H$ we see that the effective prediction horizon is narrowed for the case $P_2\Delta(-32\ fs^2)$. Instead, in the case $\Delta(-27\ fs^2)$, the horizon imposed by $K_T$ time is larger than $H$, and the effective prediction horizon remains unaffected

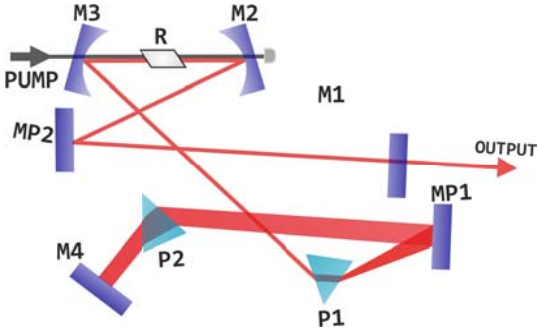

**Figure 1:** Scheme of the Ti:Sapphire laser. LB: pump focusing lens (f = 10 cm); M2,3 : curved mirrors (R = 10 cm); MP1,2 : plane HR mirrors; P1,2: pair of prisms (do not confuse with the dynamical modes of operation P1, P2). Distances in mm: M3 - R = R -M2 = 50, M2 - MP2 = 140, MP2 - M1 = 465, M3 - P1 = 297, P1 - MP1 = 198, MP1 - P2 = 415, P2 - M4 = 109. The prisms' positions are adjusted to get negative total GVD.

| Dynamical mode / Parameter | KC | $\lambda_{max}$ | $d_E$ – type of output | N |
|---|---|---|---|---|
| $P_2\ \Delta(-32)$ | 0.5280 | 0.2154 | 4 – No EE | 12004 |
| $P_2\ \Delta(-27)$ | 0.1289 | 0.1510 | 4 - EE | 12812 |
| $P_1\ \Delta(-90)$ | 0.3488 | 0.3341 | 3 – No EE | 12224 |
| $P_1 - P_2\ \Delta(-40)$ | 0.2107 | 0.2081 | 3 – No EE | 12532 |
| $P_1\ \Delta(-37)$ | 0.4965 | No measurable | No measurable – No EE | 12120 |
| $P1\ \Delta(-150)$ | 0.0458 | No measurable | periodic | 12530 |

**Table 1:** Kolmogorov complexity ($KC$) for theTi:Sapphire laser computed from experimental time series. $\Delta$: GVD in $fs^2$. $\lambda_{max}$: value of the largest positive Lyaopunov exponent. dE: embedding dimension calculated with the false nearest neighbor method. N: number of elements in the time series.

### 4. b. Nd:YVO$_4$ with slow saturable absorber.

This laser is a simple, compact, economical and robust source of nanosecond (Q-switch) laser pulses. Its application in portable rangefinders has been much studied. In our prototype, a 3×3×1mm Nd-vanadate crystal, 1% doped is longitudinally pumped by a 2W laser diode emitting at 808 nm. A standard V-shaped cavity produces a mode waist near the output (plane) mirror (Fig.2). A solid-state slow saturable absorber (Cr:YAG crystal) is placed at the waist. The mode size at the absorber is adjusted, hence changing the condition of saturation. This is the main control parameter in this system. Several dynamical regimes appear, from stable Q-switch to chaos, with and without EE. The value of $KC$ for these regimes is summarized in Table 2.

The dynamics of this system is ruled by the interplay of a few transversal modes in the laser cavity [21]. In the periodical regimes, sequences of total pulse intensities are correlated with sequences of spatial transverse patterns of the field. In the chaotic ones, a few different patterns alternate. The EEs are related with even fewer ones. The series of patterns and the pulse intensities before and after an EE are markedly repetitive. These observations demonstrate that EEs in this system follow a deterministic evolution ruled by few (nonlinearly) interacting modes.

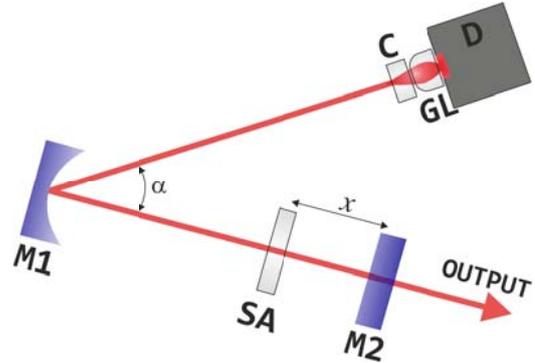

**Figure 2.** Scheme of the laser with saturable absorber. D: pump laser diode, 2W cw at 808 nm; GL: focusing GRIN lens; C: Nd:YVO$_4$ crystal active medium; M1: folding mirror (R = 100 mm); M2: output mirror (plane); SA: saturable absorber, transmission (unbleached) 90%; L: focusing lens; FPD: fast photodiode; $a$= 20°; Distance C-M1: 130 mm; Distance M1-M2: 70 mm . The position X is adjusted to obtain different dynamical regimes.

| Parameter | KC | $\lambda_{max}$ | $d_E$ - output | N |
|---|---|---|---|---|
| x=17 | 0.5980 | 0.1805 | 6 – No EE | 38941 |
| x=14 | 0.8626 | 0.2619 | 5 - EE | 30133 |
| x=8 | 0.6391 | No measurable | No measurable – No EE | 38942 |
| x=5 | 0.0461 | No measurable | periodic | 35815 |

**Table 2:** Kolmogorov complexity ($KC$) for the Nd:YVO$_4$ laser with saturable absorber computed from experimental time series. Control parameter x[mm]. $\lambda_{max}$: value of the largest positive Lyaopunov exponent. dE: embedding dimension calculated with the false nearest neighbor method. N: number of elements in the time series.

In this system, KC correctly identifies the chaotic nature of the time series, even in the case of x=8cm, where the same information isn´t enough to determine the embedding dimension and the Lyapunov exponents. Also the distiguishability between periodic and chaotic time series is greater than one order of magnitude ( from KC = 0.0461 for a periodic case to KC = 0.5980 in the less chaotic series). Unlike the Ti:Sapphire laser, here, the appearance of EE produces an increase in KC from 0.5980 in x= 17cm to 0.8626 in x= 14cm, which is coherent with the fact that we need more information to describe the extra transversal modes involved in the formation of the EE. The increase by 40% in KC along with the rise in $\lambda_{max}$ for the same amount in time series with EE, indicate that predict the value of the next pulse in presence of EE is more difficult than the same prediction in absence of EE.

### 4. c. Nd:YVO$_4$ with modulation of losses.

Haken [22] first noted the isomorphism between the equations of the single-mode laser and Lorenz equations. Yet, the parameter values for the laser case forbid the evolution of the interesting chaotic regime. In order to get an extra degree of freedom to allow chaotic evolution, Arecchi [23] proposed and developed a laser with external modulation of losses. Thanks to its reliability, high speed of data generation and richness of phenomena, this device has become a cornerstone to study nonlinear dynamics since then.

Our (all-solid-state) prototype is based on a diode-pumped Nd:YVO$_4$ crystal (as in Section 4.b, see Fig.3) with the insertion of an intra-cavity electro-optic modulator (EOM). The EOM axes are oriented at 45 degrees of the main polarization axis of the field so that it introduces a polarization rotation proportional to the applied voltage $V_{mod}$. The active medium Nd:YVO$_4$ has a strong polarization-dependent gain, with the highest gain parallel to its c-axis. The effect of the voltage applied to the EOM is hence a change in the cavity gain-losses balance. The EOM is driven at a sinusoidal frequency $f$ near the relaxation oscillations value, which is measured here 111 kHz. Depending on the values of the introduced modulation, the laser output is observed to be continuous-wave, periodically pulsed, and chaotically pulsed with, or without, EE. The value of KC for each possible type of oscillation output is summarized in Table 3.

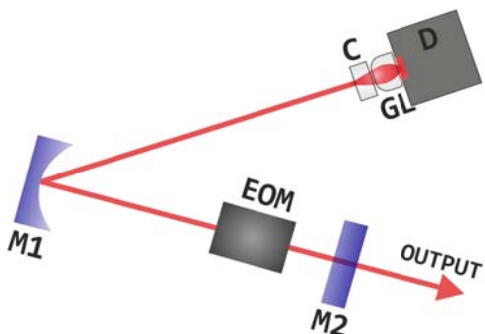

**Figure 3:** Scheme of the laser with modulation. D, pump diode; Nd:YVO$_4$ active medium; M1, HR folding mirror ROC = 25cm; M2, output coupler; R =83 %; EOM, electro - optical modulator; PD fast photodiode.

The origin and features of the dynamics displayed by this laser has been extensively studied, both theoretically and experimentally [24,25]. In few words, there are several coexisting periodic attractors for the same parameter values, implying the existence of generalized multi-stability. An increase of the modulation amplitude results in a series of abrupt expansions of the chaotic attractor yielding a kind of staircase in the bifurcation diagram. Each step marks the occurrence of a bifurcation due to the collision of a chaotic attractor with an unstable orbit. This is named an *external crisis* of the chaotic attractor. This process generates pulses whose maximum intensity are much larger than those previous to the collision, and which remain rare compared to the common of pulses (being, in consequence, EEs).

| Parameter | KC | $\lambda_{max}$ | $d_E$ - output | N |
|---|---|---|---|---|
| f=108 | 0.3973 | 0.5143 | 6 – No EE | 15803 |
| f=115 | 0.8495 | 0.7743 | 4 - EE | 14715 |
| f=106 | 0.6215 | No measurable | No measurable – No EE | 20560 |
| f=120 | 0.0372 | No measurable | periodic | 25004 |

**Table 3.** Kolmogorov complexity (KC) for the Nd:YVO$_4$ laser with modulation of losses calculated from experimental time series. Parameter $f$ [kHz] frequency of modulation. dE: embedding dimension calculated with the false nearest neighbor method. N: number of pulses of the time series.

Once again, KC is able to detect chaos (0< KC < 1) in operation regimes with ($f$ = 115kHz) and without EE ($f$ = 108kHz). It is also able to identify chaotic time series when the reconstruction method fails ($f$ = 106 kHz) and to discriminate, by a variation of one order of magnitude, against periodic time series. KC grows from 0.3973 for $f$ =108 kHz to 0.8495 for $f$ = 115 kHz when an EE appears. This increase is the consequence of the additional information (implying a longer algorithm) needed to describe the crash of the chaotic attractor with the unstable orbit of other multi - stable solution. The largest positive Lyapunov exponent also increases, from $\lambda_{max}$ = 0.5143 for $f$ = 108 kHz to 0.7743 for $f$ = 115 kHz. In consequence, the effective prediction horizon is strongly reduced for time series with EE in comparison with a time series without EE. As in the case Nd:YVO$_4$ with slow saturable absorber, this laser is more random (in the sense of less predictable) in the EE regime, despite the immediate vicinity of the EEs is more regular than for an average pulse.

## 5. Conclusion.

By using *KC* as a guide, the description of the effects of changes in the control parameter on the dynamics of the considered laser systems is greatly simplified. Because *KC* is computationally easier and faster to calculate than, say, dE, it is a practical quantity to diagnose the state of the system in a scheme of real-time control of chaos. We have shown this for experimental time series obtained from three different dynamical systems of interest. The origin of chaos and EE are different in each case, but in all cases *KC* can be reliably used to identify chaotic and EE outputs, and to get a glimpse of the underlying causes of the observed phenomena. Such a tool can then be useful, as the feedback control signal, in the development of a real time control system of the laser output.

**Acknowledgements.**

Many thanks to Prof. J.Tredicce for his observations, suggestions and encouragement. This work received support from the grants N62909-18-1-2021 Office of Naval Research Global (USA), PIDDEF 01-14 Ministerio de Defensa (Argentina) and PIP 2017 0100027C CONICET (Argentina).

**References.**

1. H.D.I. Abarbanel Analysis of Observed Chaotic Data, Springer (1996).
2. D. A. Rand and L.S. Young. Dynamical Systems and Turbulence, Lecture Notes in Mathematics, **898** pg.230 -366 (1981).
3. A. Kolmogorov Problems of Information Transmission **1** 4 (1965).
4. G. Chaitin J. of the Association for Computing Machinery **13** 547 (1966).
5. R. Solomonoff Information and Control **7** 1 (1964).
6. A. Solis, A. M. AnguloMartínez, R. RamírezAlarcón, H. Cruz Ramírez, A. B. U'Ren and J. G. Hirsch Phys. Scr**90** 074034 (2015).
7. D. Mihailovic, G.Mimic, P. Gualtieri, I.Arsenic 1 and C. GualtieriEntrophy**19** (10) 519 (2017).
8. L.Mi and P. VitanyiAn Introduction to Kolmogorov Complexity and Its Applications third edition Springer (2008).
9. A. Lempel, J. Ziv IEEE Trans. Inform. Theory **22** 75 (1976).
10. F.Kaspar and H.SchusterPhys.Rev.A**36** 842 (1987).
11. C. Kharif, E. Pelinovsky and A. Slunyaev Rogue Waves in the Ocean Springer (2009).
12. L. Draper, Mar. Obs. **35** 193 (1965).
13. D.R. Solli, C. Ropers, P. Koonath, et al., Nature **450** 1054 (2007).
14. Yu.V. Bludov, V.V. Konotop, N. Akhmediev, Phys.Rev A **80** 033610 (2009).
15. J.Eppink. Journal of Visual Culture **13** 3 298 (2014).
16. D. Mihailovic et al. Entropy, 20, 570 (2018).
17. M.G. Kovalsky and A. A. Hnilo Opt. Commun.**186** 155 (2000).
18. M. G. Kovalsky, A. A. Hnilo. Phys. Rev. A **70** 043813 (2004).
19. A. A. Hnilo, M. G. Kovalsky, M. B. Aguero, and J. R. Tredicce Phys. Rev. A **91** 013836 (2015).
20. M. G. Kovalsky, J. R. Tredicce Opt. Lett. **36** 22 (2011).
21. C. R. Bonazzola, A.A.Hnilo, M. G. Kovalsky, and J. R. Tredicce Phys. Rev. E .**97** 032215 (2018).
22. H.Haken, Phys.Lett. 53A p.77 (1975).
23. F.Arecchi, R.Meucci, G.Puccioni and J.Tredicce, Phys.Rev.Lett. 49 p.1217 (1982).
24. C. Metayer et al. Optics Express **22** 17 19850 (2014).
25. N.M. Granese et al. Opt. Lett. **41** 13 (2016).